\magnification=\magstep 1
\vsize=9.5 true in
\baselineskip=14pt
\font\bhead=cmbx10 scaled 1440
\def \d {{\rm d}}

\ \vskip1cm
\centerline{\bhead Interpreting a conformally flat}
\medskip
\centerline{\bhead pure radiation space-time}
\bigskip\bigskip

\centerline{J. B. Griffiths\footnote{$^1$}
		     {E-mail: {\tt J.B.Griffiths@Lboro.ac.uk}}} 
\smallskip
\centerline{Department of Mathematical Sciences, Loughborough University}
\centerline{Loughborough, Leics. LE11 3TU, U.K.} 

\bigskip
\centerline{and J. Podolsk\'y\footnote{$^2$}
       {E-mail: {\tt Podolsky@mbox.troja.mff.cuni.cz}}}
\smallskip
\centerline{Department of Theoretical Physics,
Faculty of Mathematics and Physics,}
\centerline{Charles University,
V Hole\v{s}ovi\v{c}k\'ach 2, 18000 Prague 8, Czech Republic.}

\medskip
\centerline{24 June 1998}

\bigskip\bigskip
\centerline{ABSTRACT}
\medskip
A physical interpretation is presented of the general class of
conformally flat pure radiation metrics that has recently been identified by
Edgar and Ludwig. It is shown that, at least in the weak field limit,
successive wave surfaces can be represented as null (half) hyperplanes rolled
around a two-dimensional null cone. In the impulsive limit, the solution
reduces to a {\sl pp}-wave whose direction of propagation depends on retarded
time. In the general case, there is a coordinate singularity which corresponds
to an envelope of the wave surfaces. The global structure is discussed and a
possible vacuum extension through the envelope is proposed. 

\vfil\noindent
PACS class 04.20.Jb, 04.30.Nk, 04.40.-b

\vfil\eject
\noindent{\bf 1. Introduction}

\smallskip
Vacuum and Einstein--Maxwell space-times of Kundt's class, which admit a
non-expanding shear-free and twist-free null geodesic congruence, are well
known [1]. A null Einstein--Maxwell field can, of course, often be
reinterpreted as some other form of pure radiation. However, Wils [2] has
found a conformally flat pure radiation space-time of this class that cannot
be interpreted as a null electromagnetic, scalar or neutrino field. This
solution was subsequently generalised by Koutras and McIntosh [3] who
expressed it in the form  
 $$ \d s^2 =2(ax+b)\d u\,\d v -2av\,\d u\,\d x
+\big[2f(u)(ax+b)(x^2+y^2)-a^2v^2\big]\d u^2 -\d x^2 -\d y^2. $$ 
 For this, the only non-zero component of the curvature tensor is
$\Phi_{22}=2f(u)/(ax+b)$, so it represents a conformally flat space-time with
null radiation. When $a=0$, $b=1$, this is just the familiar plane wave
metric. When $a=1$, $b=0$, it is the solution of Wils [2].

In fact, this still does not describe the full class of conformally flat pure
radiation metrics. The complete family has been given by Edgar and Ludwig
[4] who derived the general solution of this type as a member of Kundt's
class. In a subsequent paper [5], they re-derived the general solution using
the ${\scriptstyle{\rm GHP}}$ formalism. This general metric (excluding the
plane wave limit) can be written in the form 
 $$ \d s^2 =2x\,\d u\,\d v -2v\,\d u\,\d x 
+\big[2f(u)x\big(x^2+y^2+g(u)y+h(u)\big)-v^2\big]\,\d u^2 
-\d x^2 -\d y^2. \eqno(1) $$ 
 This can be shown to include the Koutras--McIntosh solution above (with
$a\ne0$) after a further coordinate transformation. When $g(u)=0$ and
$h(u)=0$, it is just the solution of Wils [2]. Its explicit conformal factor
was determined in [6].

This family of metrics is of particular interest to those who work on the
computer-aided classification of exact solutions. As shown in [3], in general,
they contain no invariants or Killing or homothetic vectors, and thus provide
an interesting exceptional case for classification. It has been conjectured
that any space-time can be uniquely characterized using the derivatives of the
curvature tensor in which no higher than the fourth derivative is ever
required.  An invariant classification of the above space-times has recently
been provided by Skea [7]. In this, he has shown that, to distinguish the Wils
metric within the general Edgar--Ludwig solution, it is necessary to go as far
as but no further than the fourth derivative of the curvature tensor, thus
supporting the above conjecture.

To our knowledge, a physical interpretation of these conformally flat pure
radiation metrics, and indeed of the larger class of Kundt (non-{\sl
pp})-waves, has not previously been attempted. The purpose of this paper is to
make an initial contribution towards understanding the physical properties of
these space-times. We consider here only the line element (1) which excludes
the very familiar plane-wave solutions.

\bigskip\goodbreak
\noindent{\bf 2. Initial observations}
\smallskip
It may immediately be observed from (1) that the surfaces $u=u_0=$~const. are
planes spanned by ``cartesian'' coordinates $x$ and $y$. In addition, there is
a coordinate singularity when $x=0$. In the following sections, we will
investigate the nature of this singularity and the family of successive
planes.

Since the solution contains an arbitrary ``amplitude'' function $f(u)$ which
represents a deviation from a flat background, it is appropriate to consider
the cases in which $f$ is small, or in which it describes an impulsive, step
or sandwich wave. Of course these solutions represent matter propagating with
the speed of light rather than gravitational or other forms of wave. However
the terminology of ``waves'' is natural and will be adopted, and the surfaces
$u=u_0=$~const. will be referred to as ``wave surfaces''.

\bigskip\goodbreak
\noindent{\bf 3. The solution in NP notation} 
\smallskip

The metric (1) can be expressed in the contravariant form 
 $$ g^{\mu\nu}=\pmatrix{0 &{1\over x} &0 &0 \cr \noalign{\medskip}
{1\over x} &-{2\over x}H &-{v\over x} &0 \cr \noalign{\medskip}
0 &-{v\over x} &-1 &0 \cr \noalign{\medskip}
0 &0 &0 &-1 \cr } \eqno(2) $$ 
 where $(x^1,x^2,x^3,x^4)=(u,v,x,y)$ and 
$H(u,x,y)=f(u)\big(x^2+y^2+g(u)y+h(u)\big)$. Using the Newman--Penrose
formalism, a tetrad can be chosen such that 
 $$ D=\partial_v, \qquad 
\Delta={1\over x}\big(\partial_u-H\partial_v-v\partial_x\big), \qquad
\delta={1\over\sqrt2}\big(\partial_x-i\partial_y\big). \eqno(3) $$ 
 The only non-zero ${\scriptstyle{\rm NP}}$ quantities are then 
 $$ \tau=-{1\over\sqrt2\,x}, \qquad
\nu={f(u)\over\sqrt2\,x} \big[2(x+iy)+ig(u)\big], \qquad 
\Phi_{22}={2f(u)\over x}. \eqno(4) $$ 
 Since $\kappa=\rho=\sigma=0$, it follows that the congruence tangent to the
null radiation is geodesic with zero expansion, twist and shear (i.e. the
metric is of Kundt's class). The tetrad that has been chosen is parallelly
propagated along the congruence. However, the fact that $\tau$ is nonzero
indicates that the congruence rotates about some direction perpendicular to
that in which it propagates.

Since $\Phi_{22}$ is the only non-zero component of the curvature tensor, the
scalar invariants all vanish, and so $x=0$ cannot correspond to a scalar
curvature singularity. Moreover, on rescaling the tetrad by putting
$\tilde\ell^\mu=(1/\sqrt x)\,\ell^\mu$, $\tilde n^\mu=\sqrt x\,n^\mu$, the
non-zero Ricci tensor component is given by $\tilde\Phi_{22}=2f(u)$. If $f$
is non-zero everywhere, a further rescaling can be used to make it equal to
unity, but since we wish to include the possibilities of impulsive, step or
sandwich waves, we will here retain $f(u)$ as an arbitrary amplitude function.

\bigskip\goodbreak
\noindent{\bf 4. The source of the solution} 
\smallskip

Let us now consider the possible sources of these space-times. These contain
pure radiation, which can always be considered phenomenologically as a ``null
fluid'' or ``null dust''. However, pure radiation solutions can sometimes be
interpreted as representing a null electromagnetic field, a massless scalar
field or a neutrino field. In fact, none of these interpretations are possible
for this class of space-times. To demonstrate this, we first note that the
only non-zero component of the Ricci tensor is $\Phi_{22}$. In this case, the
Einstein--Maxwell equations can only be satisfied if a function $\Phi_2$ can
be found such that 
 $$ D\Phi_2=(\rho-2\epsilon)\Phi_2, \qquad \delta\Phi_2=(\tau-2\beta)\Phi_2,
\qquad 8\pi\Phi_2\bar\Phi_2=\Phi_{22}. $$ 
 Alternatively, the space-time will admit a massless scalar field $\varphi$ if 
 $$ D\varphi=\delta\varphi=\bar\delta\varphi=0, \qquad
\Delta\varphi\Delta\bar\varphi=\Phi_{22}. $$ 
 Finally, the space-time will admit a (two-component) massless neutrino
interpretation if a function $\phi$ can be found such that 
 $$ D\phi=(\rho-\epsilon)\phi, \quad \delta\phi=(\tau-\beta)\phi, \quad
\bar\delta\phi =(2\bar\tau-\alpha)\phi, 
\quad 8\pi i\big[\phi\Delta\bar\phi -\bar\phi\Delta\phi
+(\bar\gamma-\gamma)\phi\bar\phi\big] =\Phi_{22}. $$ 
 With the non-zero NP quantities (4), it can be shown that none of these sets
of equations can be satisfied. This is consistent with the fact that the Wils
metric [2] was demonstrated not to be a null electromagnetic, scalar or
neutrino field.

We must thus conclude that these space-times represent some form of ``null
fluid'' which must be composed of some form of incoherent null radiation.

\bigskip\goodbreak
\noindent{\bf 5. Wave surfaces in the weak field limit} 
\smallskip

Let us now investigate the cases in which the arbitrary ``amplitude'' function
$f(u)$ is small, or in which it represents an impulsive, step or sandwich
wave. In the latter cases, $f(u)$ is zero ahead of the wave where the
space-time is flat and the line element (1) reduces to the form  
 $$ \d s^2 =2x\,\d u\,\d v -2v\,\d u\,\d x -v^2\,\d u^2 -\d x^2 -\d y^2.
\eqno(6) $$ 
 This is the background into which the step or impulsive wave propagates. To
determine the geometry of the wavefront, we need to express this in standard
cartesian coordinates and then describe the front $u=u_0=$~const. in these
coordinates. This would give a clear geometrical and physical interpretation. 
When considering the alternative case in which $f$ is small, the metric (1)
represents a small perturbation of Minkowski space, and the wave surfaces can
similarly be described by considering the null hypersurfaces $u=u_0$ in
cartesian coordinates.

The necessary step can be achieved using the transformation 
 $$ \eqalign{ U&=(2x+uv)u \cr
V&=v \cr
X&=x+uv \cr
Y&=y \cr } \qquad\qquad \Leftrightarrow \qquad\qquad
\eqalign{u&=\big(X\mp\sqrt{X^2-UV}\big)/V \cr
v&=V \cr
x&=\pm\sqrt{X^2-UV} \cr
y&=Y. \cr } \eqno(7) $$
 This takes the background (Minkowski) part (6) of the metric (1) to the
obviously flat form 
 $$ \d s^2=\d U\,\d V -\d X^2-\d Y^2. $$ 
 The surfaces $x=$~const. can be seen, in this background, to be
hyperboloidal, degenerating to a 2-dimensional null cone $X^2=UV$ when $x=0$.

It can then be seen that the null surface $u=u_0=$ const. is given by 
 $$ (1+u_0^2)T-2u_0X-(1-u_0^2)Z=0, \eqno(8) $$ 
 where $U=T-Z$ and $V=T+Z$. These surfaces are a family of null hyperplanes
whose orientation varies for different values of $u_0$. Putting 
 $$ \sin\alpha={2u_0\over1+u_0^2}, \qquad {\rm or \ equivalently} \qquad
u_0=\tan{\alpha\over2}, \eqno(9) $$ 
 the corresponding null normal $N^\mu$ is given using coordinates $(T,X,Y,Z)$
by
 $$ N^\mu=\big(1,\sin\alpha ,0,\cos\alpha \big). $$ 
 This clearly demonstrates that, at any time, successive wave surfaces 
$u=u_0$ are rotated about the $Y$ axis as $u_0$ increases from $-\infty$ to
$+\infty$ (or as $\alpha$ goes from $-\pi$ to $+\pi$). The rotation of these
planes for different values of $u_0$ is consistent with the non-zero value of
$\tau$ for these metrics.

This basic geometrical structure of the background enables us to interpret the
behaviour of impulsive, shock or sandwich waves in which the profile function
$f(u)$ takes appropriate forms. In these cases $f$ vanishes for all $u$ up to
some constant $u_0$ representing the wavefront. The particular value of $u_0$
determines the orientation of the initial wavefront in the Minkowski
background. Alternatively, when $f$ is small, successive wave surfaces can be
seen to rotate relative to the approximate Minkowski background.

\bigskip\goodbreak
\noindent{\bf 6. Global structure in the weak field limit} 
\smallskip

{}From the transformation (7) it is obvious that, in the weak field limit, the
coordinate singularity at $x=0$ occurs on the hypersurface \ $T^2=X^2+Z^2$ \
in Minkowski space-time. On suppressing the $Y$ coordinate, which may be
arbitrary, this is a two-dimensional null cone. In 3-dimensional space, it is
a cylinder with radius expanding at the speed of light for $T>0$. (For a
complete space-time, we must also include the case when $T<0$ in which the
cylinder is contracting.)

The plane wave surfaces $u=u_0$ are spanned by $x$ and $y$. With (9), these
hyperplanes (8) are given by \ $T-\sin\alpha\,X-\cos\alpha\,Z=0$ \ and are
tangent to the null cone or expanding cylinder. For successive wave surfaces
with different values of \ $\alpha=2\arctan u_0$, the corresponding null
hyperplane rolls around the surface of this 2-dimensional null cone as
$\alpha$ increases from $-\pi$ to $\pi$ (i.e. $u_0$ goes from $-\infty$ to
$\infty$).

Of course, the metric (1) is strictly only a local solution of Einstein's
equations. Naturally, we want to interpret it over as wide a range of
coordinates as possible. With the above interpretation, from any point in
space at any time $T$, there are two planes that are tangent to the expanding
null cylinder. But the pure radiation is always directed perpendicular to the
planes and, if two distinct plane wave surfaces pass through a point, there
will exist radiation with two null components. This is clearly ruled out by
our initial assumptions (only $\Phi_{22}$ non-zero). Thus, the plane wave
surfaces must be restricted in some way.

In fact, such a restriction is automatically achieved by imposing the
condition that $x\ge0$. From (7) we see that, on a given wave surface $u=u_0$,
we have $X-u_0(T+Z)=x$. Substituting $T$ using (8) gives 
 $$ \cos\alpha\,X-\sin\alpha\,Z=x. \eqno(10) $$ 
 This is valid on any surface at any time $T$. Imposing the natural condition
that $x\ge0$ effectively restricts the wave surface to a half of each null
plane.

{}From now on a wave surface will be considered as the half-plane
satisfying (8) with this restriction. In this case, for any point outside the
cone at any time there exists only one wave surface
$u=u_0\in(-\infty,\infty)$ passing through it, thus resolving the physical
problem mentioned above. The entire space outside the null cone is now
covered exactly once. The wave surfaces for $T$ and $Y$ constant are
illustrated in figure~1. The restriction $x\in[0,\infty)$ also resolves the
mathematical ambiguity in the signs in the right hand side of (7) in which
upper signs only are now assumed. This ambiguity corresponds to the invariance
of the metric (6) under the reflections $x\to-x$, $v\to-v$. To retain this
invariance in the full metric (1) we can also transform $f\to-f$ as is
required to maintain the sign of $\Phi_{22}$.

The wave surface (8), $u=u_0$ with $x\ge0$, is tangent to the null cone or
expanding cylinder as illustrated in figure~2. Any half-plane wave surface is
bounded by the line $x=0$ where it touches the null cone, i.e. on
$X=\tan\alpha\,Z$, $Y$ arbitrary. For successive wave surfaces with different
values of \ $\alpha=2\arctan u_0$, the corresponding null half-hyperplane is
rotated around the surface of this 2-dimensional null cone.

This analysis of successive wave surfaces as hyperplanes rolling over a cone
with only half of each plane considered is applicable in the weak field limit
also to the family of (non-{\sl pp}-wave) type~N solutions of Kundt's
class\footnote{$^1$}
    {Note added in proof: For a general discussion of constructions of this
     type see [10].}.  
     For the conformally flat solutions, this geometrical construction does
not distinguish between the Wils solution and the more general Edgar--Ludwig
solution.

\bigskip\goodbreak
\noindent{\bf 7. The impulsive limit} 
\smallskip

Let us now consider the case of an impulsive wave representing a thin shell of
null matter on an arbitrary wave surface $u=u_0$ in the space-time
described by (1). Putting $f(u)=c\delta(u-u_0)$ and using the transformation
(7), we obtain 
 $$ \eqalign { \d s^2&=\d T^2 -\d X^2 -\d Y^2 -\d Z^2 \cr 
&\qquad\qquad +{\cal A}\>\delta(T-\sin\alpha\,X-\cos\alpha\,Z) 
(\d T-\sin\alpha\,\d X-\cos\alpha\,\d Z)^2 \cr } \eqno(11) $$ 
 where the amplitude function ${\cal A}$ is given by 
 $$ {\cal A}= c(1+u_0^2) \big[ (\cos\alpha\,X-\sin\alpha\,Z)^2+Y^2 +g(u_0)Y
+h(u_0) \big]  $$ 
 in which $u_0$ is related to $\alpha$ by (9).

In this particular case, there is no ambiguity in extending the wave surface
$u=u_0$ to the entire hyperplane (8). The restriction $x\ge0$ need not be
applied. However, it may be recalled that $\Phi_{22}=2f(u)/x$, so the natural
extension results in a negative energy density over half of the extended wave
surface. An alternative extension in which the additional half-plane is simply
a reflection of the original so that $\Phi_{22}=2f(u)/|x|$ may also be
considered.

The resulting complete metric (11) is clearly an impulsive {\sl pp}-wave
propagating in the direction \ $\sin\alpha\,X+\cos\alpha\,Z$ \ corresponding
to $N^\mu$ in a Minkowski background. This reduction to the familiar {\sl
pp}-wave is as expected since, at any time, the wavefront is plane. It may
also be noted that this reduction is similar to the reduction of vacuum type~N
Kundt waves to {\sl pp}-waves in the impulsive limit~[8].

Comparing the family of such impulsive solutions having different values for
$u_0$, the relative rotations of the different wave surfaces is clearly
illustrated. For example, when $u_0=0$, the impulse propagates in the $Z$
direction while, for $u_0=\pm1$, it propagates in the positive or negative $X$
directions. It can further be seen that the two additional arbitrary functions
$g(u)$ and $h(u)$ play the role of free parameters in the amplitude function
${\cal A}$ on different wave surfaces. Although these may be removed by a
coordinate transformation on any particular null hypersurface $u=u_0$, they
represent a non-trivial freedom in the full solution given by (1).

\bigskip\goodbreak
\noindent{\bf 8. The coordinate singularity} 
\smallskip

We now consider the character of the singularity at $x=0$ in more detail. This
is clearly not a curvature singularity as the polynomial curvature scalars all
vanish. According to the classification scheme of Ellis and Schmidt [9], it
may be described as a quasi-regular singularity. In the weak field limit, we
have seen that the singular surface $x=0$ is a 2-dimensional null cone. It is
evident that it can be understood as an envelope of the wave surfaces (see
figure~1).

The fact that this is merely a coordinate singularity may be confirmed by
noting that on applying the transformation (7) to the exact line element (1),
we obtain 
 $$ \d s^2=\d T^2 -\d X^2 -\d Y^2 -\d Z^2 
 +{\cal B}\> \big[(1+u^2)\d T-2u\,\d X-(1-u^2)\d Z\big]^2 \eqno(12) $$ 
 where 
 $$ u(T,X,Z)={X-\sqrt{X^2+Z^2-T^2}\over T+Z} $$ 
 and 
 $$ {\cal B}=f(u){X^2+Y^2+Z^2-T^2+g(u)Y+h(u)\over2\sqrt{X^2+Z^2-T^2}}. $$ 
 It may be observed that the line element (12) is exactly of the Kerr--Schild
form, and that this is non-singular (det$\,g_{\mu\nu}=-1$) everywhere even on
$x=\sqrt{X^2+Z^2-T^2}=0$. It can also be seen that (12) reduces to (11) in the
impulsive limit.

\bigskip\goodbreak
\noindent{\bf 9. A possible extension of the space-time} 
\smallskip

In interpreting the solution (1) we made the restriction $x\ge0$ with upper
signs taken in (7). This was necessary for its physical interpretation
as a pure radiation field. This restricted the wave surfaces to a family of
half planes. In this way, we have interpreted a ``local'' solution up to its
natural boundary on the coordinate singularity at $x=0$. We now consider the
possibility of extending the solution through this hypersurface.

It has been argued that the coordinate singularity arises as an envelope of
the null hyperplanes rolling around a 2-dimensional null cone, and that the
metric (1) applies only to the exterior region. Moreover, the envelope $x=0$
is a null hypersurface. In these circumstances, it is possible to include the
appropriate section of Minkowski space inside the cone. In this case, a
complete solution (for $T>0$) can be constructed which represents outgoing
null matter leaving a cylindrical void in the region $X^2+Z^2<T^2$. Such an
extension, of course, is not unique.

\bigskip\goodbreak
\noindent{\bf 10. Further remarks} 
\smallskip

Since the coordinate range in (1) has been restricted such that
$x\in[0,\infty)$, and since $\Phi_{22}=2f(u)/x$, to ensure that the energy
density is non-negative everywhere it is necessary that the arbitrary function
$f(u)$ is chosen such that $f(u)\ge0$.

Let us also consider how the radiation would appear to an observer at a fixed
point $(X,Y,Z)$ in space-time. Assuming that this is initially ($T=T_0>0$) way
outside the null cone and that $f>0$ for all $u$, the radiation would appear
to come from a direction perpendicular to the null hyperplane $u=u_0$ which is
tangent to the boundary of a cylindrical void of radius $T_0$ centred at the
origin and which passes through $(X,Y,Z)$. Then, as $T$ increases, the
direction of the radiation will rotate. This will continue until the boundary
reaches the observer at which point the direction of the radiation is exactly
radial. After this event, the radiation will cease (provided the space-time
has been extended with an interior void).

Of course the magnitude of the radiation detected by the observer will depend
on the arbitrary ``amplitude'' function $f(u)$. The other arbitrary functions
$g(u)$ and $h(u)$ cannot be measured directly as they do not appear in the
Ricci tensor and there is no gravitational radiation. However, they cannot be
removed by a coordinate transformation unless they are constants. To detect
the effect of these terms, it would be necessary to employ a family of
observers equipped to determine the higher derivatives of the curvature
tensor. As shown by Skea [7], the functions $g(u)$ and $h(u)$ appear
explicitly in the second derivatives of the curvature tensor, although the
fourth derivatives are required to distinguish between the Edgar--Ludwig
solutions with $g$ and $h$ nonconstant and the Wils subclass.

It may finally be observed that pure radiation solutions which can be
expressed in the Kerr--Schild form have been discussed in [1]. In fact, the
solution (1) is implicitly contained in the general form given in \S28.4.3 of
[1] for type~N pure radiation solutions. However, the conformally flat case
described here and in [4--5] is not given explicitly.

\bigskip\bigskip\goodbreak
\noindent{\bf Acknowledgments}

This work was supported by a visiting fellowship from the Royal Society and,
in part, by the grant GACR-202/96/0206 of the Czech Republic and the grant
GAUK-230/96 of Charles University.

\bigskip\bigskip\goodbreak
\noindent{\bf References}
\medskip

\item{[1]} Kramer D, Stephani H, MacCallum and Herlt E 1980  {\it Exact
solutions of Einstein's field equations} (Cambridge: Cambridge University
Press) 

\item{[2]} Wils P 1989 {\sl Class. Quantum Grav.} {\bf 6} 1243

\item{[3]} Koutras A and McIntosh C 1996 {\sl Class. Quantum Grav.} {\bf 13}
L47

\item{[4]} Edgar S B and Ludwig G 1997 {\sl Class. Quantum Grav.} {\bf 14} L65

\item{[5]} Edgar S B and Ludwig G 1997 {\sl Gen. Rel. Grav.} {\bf 29} 1309 

\item{[6]} Ludwig G and Edgar S B 1997 {\sl Class. Quantum Grav.} {\bf 14}
3453 

\item{[7]} Skea J E F 1997 {\sl Class. Quantum Grav.} {\bf 14} 2393 

\item{[8]} Podolsk\'y J 1998 {\sl Class. Quantum Grav.} {\bf 15} to appear 

\item{[9]} Ellis G F R and Schmidt B G 1977 {\sl Gen. Rel. Grav.} {\bf 8} 915 

\item{[10]} Urbantke H 1979 {\sl J. Math. Phys.} {\bf 20} 1851

\bigskip\bigskip\goodbreak
\noindent{\bf Captions for figures}
\medskip

\noindent {\bf Figure 1}

For constant $T$ and $Y$, the wave surfaces $u=u_0=$~const. in the weak field
limit can be represented as a family of half lines at a perpendicular distance
$T$ from an origin as indicated. The envelope of the lines forms a circle
corresponding to the coordinate singularity at $x=0$. As $T$ increases, the
circle expands and the null matter on each wave surface propagates
perpendicular to that surface. 

\bigskip

\noindent {\bf Figure 2}

In space-time, the half-plane wave surfaces are tangent to a 2-dimensional
null cone on which $x=0$. The pictures represent two sections at constant $Y$
and at constant $T$.

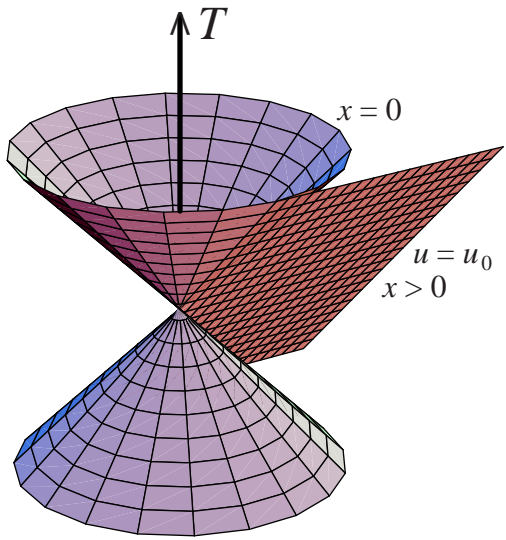
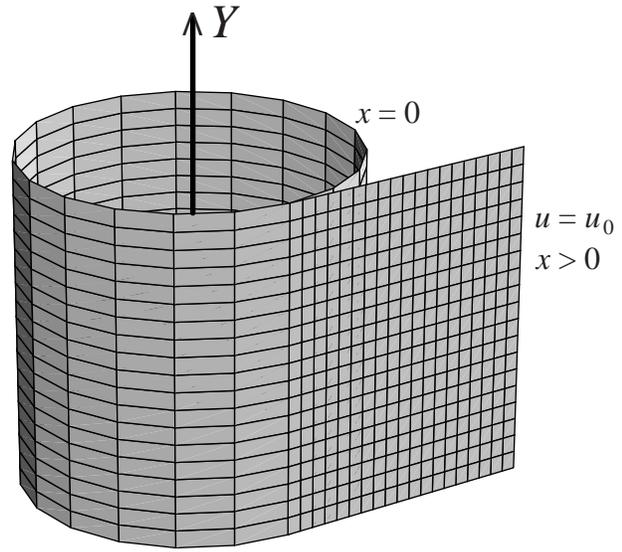
\bye